\documentclass[journal=jpccck,manuscript=article,layout=twocolumn,articletitle=true]{achemso}

\makeatletter
\renewcommand*\titlesize{\Large}
\SectionNumbersOn
\makeatother

\usepackage[utf8]{inputenc}
\usepackage[T1]{fontenc}
\usepackage[english]{babel}

\usepackage{amsmath, amsfonts, amssymb}
\usepackage{graphicx}
\usepackage[svgnames]{xcolor}
\usepackage{hyperref}
\usepackage{multirow}
\usepackage{fancyhdr}

\newcommand\e[1]{\textsubscript{#1}}
\newcommand\ex[1]{\textsuperscript{#1}}

\fancypagestyle{plain}{
\fancyhf{}

\fancyhead[C]{~\\ \textsf{\href{http://doi.org/10.1021/acs.jpcc.6b08594}{\textcolor{blue}{Published as: \emph{J. Phys. Chem. C} \textbf{2016}, 120, 24885–24894,\\ DOI: 10.1021/acs.jpcc.6b08594}}}}
}
\title{Heterometallic Metal--Organic Frameworks of MOF-5 and UiO-66 Families: Insight from Computational Chemistry}

\author{Fabien Trousselet}
\affiliation{Chimie ParisTech, PSL Research University, CNRS, Institut de Recherche de Chimie Paris, 75005 Paris, France}
\author{Aur\'elien Archereau}
\affiliation{Chimie ParisTech, PSL Research University, CNRS, Institut de Recherche de Chimie Paris, 75005 Paris, France}
\alsoaffiliation{\'Ecole Normale Sup\'erieure, PSL Research University, D\'epartement de Chimie, 24 rue Lhomond, 75005 Paris, France}
\author{Anne Boutin}
\affiliation{\'Ecole Normale Sup\'erieure, PSL Research University, D\'epartement de Chimie, Sorbonne Universit\'es -- UPMC Univ Paris 06, CNRS UMR 8640 PASTEUR, 24 rue Lhomond, 75005 Paris, France}
\author{Fran\c{c}ois-Xavier Coudert}
\email{fx.coudert@chimie-paristech.fr}
\affiliation{Chimie ParisTech, PSL Research University, CNRS, Institut de Recherche de Chimie Paris, 75005 Paris, France}

\begin{document}

\begin{abstract}

We study the energetic stability and structural features of bimetallic metal--organic frameworks. Such heterometallic MOFs, which can result from partial substitutions between two types of cations, can have specific physical or chemical properties used for example in catalysis or gas adsorption. We work here to provide through computational chemistry a microscopic understanding of bimetallic MOFs and the distribution of cations within their structure. We develop a methodology based on a systematic study of possible cation distributions at all cation ratios by means of quantum chemistry calculations at the density functional theory level. We analyze the energies of the resulting bimetallic frameworks and correlate them with various disorder descriptors (functions of the bimetallic framework topology, regardless of exact atomic positions). We apply our methodology to two families of MOFs known for heterometallicity: MOF-5 (with divalent metal ions) and UiO-66 (with tetravalent metal ions). We observe that bimetallicity is overall more favorable for pairs of cations with sizes very close to each other, owing to a charge transfer mechanism inside secondary building units.  For cations pairs with significant mutual size difference, metal mixing is globally less favorable; and the energy signifantly correlates with the coordination environment of linkers, determining their ability to adapt the mixing-induced strains. This effect is particularly strong in the UiO-66 family, because of high cluster coordination number.

\end{abstract}

\maketitle

\section{Introduction}

Metal--organic frameworks (MOFs) are a class of nanoporous materials constructed in a modular approach by the combination of inorganic nodes and organic linkers.\cite{Furukawa2013} They have shown great promise for gas storage and separation applications, catalysis, and drug delivery. Their design, structure, and properties can be varied by modification of the organic linkers, which can have different lengths, topologies, and geometries and can incorporate functional groups, for example to enhance preferential binding of guest substrates via optimized pore shapes/diameters for molecular separation. Their topology, chemical stability, and catalytic properties can also be tuned by modifying the nature of the coordination bonds involved.

One of the recent directions in MOF research has been the drive toward creating multifunctional MOFs (sometimes also called ``smart'' MOFs), by incorporating several functions in a single material: either \emph{multivariate MOFs} or \emph{heterogeneous MOFs} encompassing several functions by the cumulation of different chemical groups or active sites with different activities;\cite{Furukawa2015} or \emph{stimuli-responsive MOFs} that respond to external stimulus by a change in their chemical or physical properties, developing new activity under stimulation.\cite{Coudert2015} The most natural avenue for multivariate MOFs is to incorporate a large number of different functionalities on their organic linkers, by mixing functionalized linkers based on the same backbone and bearing different chemical groups. Multivariate MOFs based on MOF-5 have been demonstrated that can contain up to eight distinct functionalities in one phase, with ordered framework backbone but disordered functional groups.\cite{Deng2010} The combination of a variety of functionalized linkers can also in some cases give rise to novel and complex topologies, as was shown in the case of multivariate MOF-177.\cite{Zhang2015}

Another possibility to obtain multifunctional MOFs is to design \emph{heterometallic MOFs} (or mixed-metal MOFs), with different metal cations in the inorganic clusters of the MOF. Relatively simple bimetallic MOFs have been reported early in the advancement of MOF research, usually combining a pre-formed coordination complex (acting as a secondary building unit, or SBU) with a metal salt to build up a three-dimensional framework.\cite{He2005, Halper2005, Murphy2005} However, more complex heterometallic MOFs containing larger numbers of cations, or two types of SBUs,\cite{Zou16} have only be recently reported. One striking example is that of Yaghi's family mixed-metal MOF-74, which are microcrystalline MOF-74 frameworks with up to 10 different kinds of divalent metals (Mg, Ca, Sr, Ba, Mn, Fe, Co, Ni, Zn, and Cd) incorporated into the structure.\cite{Wang2014} These heterometallic MOFs can present an impact on the performances of the material, due to the addition of the functions of their metal centers or through synergistic effects of the heterometals. This was demonstrated, for example, for CO\e{2} capture in the CPM-200 family of materials.\cite{Zhai2016}

Yet despite this interest in heterometallic MOFs for their performance in applications, a detailed description and characterization of these heterometallic systems is still superficial --- as with disordered complex solids. Experimentally, access to the exact composition of the material can be obtained, but information on the distribution of metals is not available. It is to be noted that even in the case of multifunctional MOFs, the mapping of the different functional groups (random, well-mixed, or clustered) is very difficult to determine.\cite{Kong2013} Yet, it is of particular importance: in solid state science, the existence of correlated disorder drastically affects materials physical and chemical properties, and is key to a wide range of useful functionalities. In the MOF area, there is a growing realization that such might be the case too.\cite{Cheetham2016} It was recently demonstrated that correlated disorder --- the presence of complex states arising from the distribution of species within a crystalline material --- is present in UiO-66(Hf) with linker vacancies.\cite{Cliffe2014} While this correlated disorder can be quite hard to evidence, it strongly affects the physicochemical properties of a material.\cite{Cliffe2015}

Theoretical and computational chemistry tools have been widely used in the understanding of disorder in solid state physics in general and inorganic materials in particular. However, there have been few computational studies of heterometallic MOFs, and all of the published works, to our knowledge, assume perfectly disordered metal cations and focus on the impact of the mixed metals for specific properties such as adsorption. Such an example is the study by Lau et al. of the impact of post-synthetic exchange of Zr by Ti in UiO-66(Zr) on carbon dioxide adsorption.\cite{Lau2013} Here, we focus on describing a computational methodology, based on quantum chemical calculations at the Density Functional Theory (DFT) level, for the study of heterometallic MOFs. The methodology allows us to predict whether, for a given combination of metal centers, one can expect random distribution of the cations or clustering; as well as to understand which physical/chemical features have a dominant impact on the energy and why some specific substitution patterns are preferred. We showcase it on two archetypical families of MOFs, namely MOF-5 and UiO-66, and show how some simple chemical reasoning can explain the trends observed.

\section{Methods used\label{sec:methods}}

\subsection{Density Functional Theory calculations}

In this work we study two families of bimetallic MOF systems, where each cation site is occupied by one of two metal atoms. To model them, we use quantum chemistry calculations at the Density Functional Theory (DFT) level, with the CRYSTAL14 software package.\cite{CRYSTAL} It describes fully periodic structures, uses localized atom-centered basis sets and takes advantage of symmetry of the crystal structures. As such, it is well suited to the porous MOFs subject of this study. The basis sets we chose can be found in the software's basis set online library, and below we give the corresponding acronyms in this library:

\noindent
C: \texttt{C\_6-31d1G\_gatti\_1994} \cite{Gatti1994}\\
H: \texttt{H\_3-1p1G\_1994} \cite{Gatti1994}\\
O (in MOF-5): \texttt{O\_6-31d1\_gatti\_1994} \cite{Gatti1994}\\
O (in UiO-66), Zr: basis sets used in Ref.~\cite{Valenzano2011}\\
Ti : \texttt{Ti\_86-411(d31)G\_darco\_unpub} \cite{Cora2005}\\
Hf : \texttt{Hf\_ECP\_Stevens\_411d31G\_munoz\_2007} (pseudopotential) \cite{MunozRamo2007}\\
Ce : \texttt{Ce\_ECP\_Meyer\_2009} (pseudopotential) \cite{Graciani2011}\\
Be : \texttt{Be\_6-211d1G\_2012}\cite{Baima2013}\\
Mg : \texttt{Mg\_8-511d1G\_valenzano\_2006} \cite{Valenzano2007}\\
Ca : \texttt{Ca\_86-511d21G\_valenzano\_2006} \cite{Valenzano2006}\\
Zn : \texttt{Zn\_86-411d31G\_jaffe\_1993} \cite{Jaffe1993}\\
Cd : \texttt{Cd\_dou\_1998} \cite{Dou1998}\\
Sr : \texttt{Sr\_HAYWSC-311(1d)G\_piskunov\_2004} \cite{Piskunov2004}\\
Ba : \texttt{Ba\_HAYWSC-311(1d)G\_piskunov\_2004} \cite{Piskunov2004}

For each of the two frameworks, the exchange-correlation functional was chosen among several candidates (at the Generalized Gradient Approximation level, hybrid or not) to ensure a good agreement with experimental data (e.g. cell parameters, metal-oxygen coordination distances) on reference structures MOF-5(Zn) and UiO-66(Zr). The chosen functionals were B3LYP\cite{B3LYP} for MOF-5 structures (which has been well validated in the published literature\cite{Civalleri2006}), and PBESOL0\cite{Perdew2009} for UiO-66 structures (which gives good agreement with known experimental data, see Table~S2). The use of Grimme-type dispersion corrections\cite{Grimme2006} was originally tested, but as the bimetallic structures studied here are all of similar density and intermolecular distances, the effect of the corrections was found to be insignificant and results reported in this manuscript are thus obtained without dispersion corrections.

Reciprocal space sampling were carried out with a \textbf{k}-point mesh generated using the Monkhorst-Pack method \cite{Monkhorst1976}. Given the large sizes of unit cells, a $1\times 1\times 1$ mesh (sampling limited to the $\Gamma$ point) was used in all cases, except for bimetallic UiO-66 samples from substitutions in a conventional cell, where structure-dependent meshs were used (e.g. $2\times 2\times 1$ or $2\times 2\times 2$, depending on the cell shape) to ensure high accuracy results.

Geometry optimizations were performed with the standard updating scheme in CRYSTAL14; standard convergence criteria (maximally 0.0012~a.u. on atomic displacements during one optimization step, and 0.0003~a.u. on forces) in the MOF-5 case, while for UiO-66 higher convergence criteria were used (0.0005~a.u. and 0.0001~a.u. on displacements and forces respectively). Input files and DFT-optimized structures are available in the online repository at \url{https://github.com/fxcoudert/citable-data}.

\subsection{Study of bimetallic structures}

The procedure we use to design bimetallic structures relies on CRYSTAL14's tools for the description of disordered solids and solid solutions,\cite{Mustapha2013} which has been applied in previous work to describe binary inorganic solids, such as binary carbonates with calcite structure or the binary spinel solid solution Mg(Al,Fe)\e{2}O\e{4}.\cite{DArco2013} We describe it briefly below, and illustrate it in the case of MOF-5:
\begin{enumerate}\renewcommand{\theenumi}{\alph{enumi}}
\item select a reference cell of the MOF studied (e.g. the primitive cell of MOF-5, containing 8 cation sites);
\item list all possible cation substitutions within this cell, starting from a homometallic one (2$^8$ possibilities in the example);
\item among these structures, identify symmetry-related equivalent structures and retain only one, ending up with $p$ distinct bimetallic structures in addition to the 2 homometallic structures (in the MOF-5 case, $p=20$);
\item for each structure, determine the remaining symmetry --- the space group in each case is a subgroup of the original homometallic framework ($Fm\bar{3}m$ for MOF-5);
\item for each of the $p$ structures, perform a full energy minimization (optimizing both unit cell parameters and atomic positions) within its own space group.
\end{enumerate}
The number of structures thus generated (and thus the computational effort) grows exponentially with the size of the reference cell chosen, which can be the primitive cell, conventional cell or even a supercell --- which will be useful in the case of UiO-66 (Section~\ref{sec:UiO}).

We then analyze the relative stability of bimetallic structures with respect to homometallic ones, via their \emph{mixing energies}. For a substitution pattern labeled $j$ ($1 \leq j \leq p$) with a substitution rate $x_j\in\ ]0;1[$ of element A by element B, if the energy after relaxation is E$^{(j)}$, we define the mixing energy as: \begin{equation} E_m^{(j)} = E^{(j)} - x_jE\e{B} - (1-x_j)E\e{A} \end{equation} where $E\e{A}$ and $E\e{B}$ are the energies of the homometallic frameworks. A mixing energy $E_m<0$ indicates that a crystal with this pattern is energetically stable with respect to demixing into A- and B-based MOFs (at $T=0$ and $P=0$).

In addition to the mixing energies, we also extract from the DFT calculations other properties of bimetallic structures: topology descriptors depending on the substitution pattern's topology, rather than on the exact atomic positions after relaxation; coordination distances between cations and carboxylate oxygens (averaged spatially on the relaxed structure); and the distribution of electronic charge, measured by the Mulliken partial atomic charges on cations in the relaxed structure's ground state.

Finally, we also consider isolated clusters centered on metal nodes, formed by replacing bridging ligands by non-bridging formate groups. Mixing energies $E_m$, defined as above and obtained from relaxing such 0-D systems, reflect more directly the local effects governing the mixing, independently from lattice effects. We relaxed various clusters of type A$_{n-1}$B$_1$O, where $n=4$ (MOF-5) or $n=6$ (UiO-66) and A, B are two metal elements.

\section{Heterometallic derivatives of MOF-5\label{MOF5}}

\begin{figure}[t]
\begin{center}
\includegraphics[width=84mm]{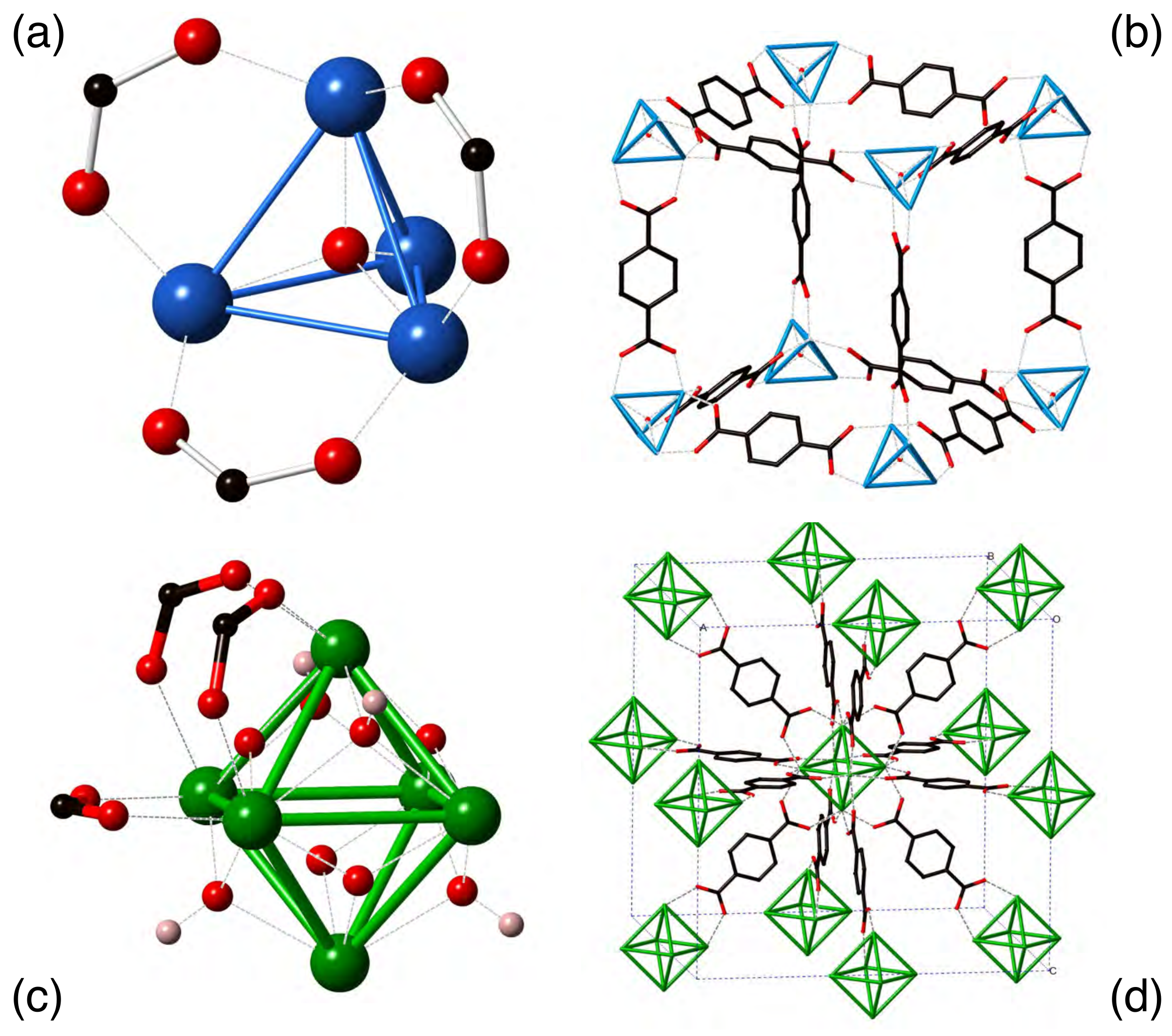}
\caption{\label{2clus2cell} (a,b) MOF-5(Zn) and (c,d) UiO-66(Zr) structures, considered in this study. (a,c) show the corresponding structure's individual SBU, plus 3 of the carboxylate groups coordinating it; (b,d) depict the corresponding framework, showing all SBUs from a conventional unit cell plus (some) linkers connecting them.  In (b) and (d), O atoms from SBUs and H atoms are not shown for clarity. Red, Black, Green and Blue stand for O, C, Zr, and Zn respectively. Covalent and metal-oxygen bonds are shown by continuous and dashed lines respectively; fictitious metal-metal bonds are also shown to highlight the cluster shape.}
\end{center}
\end{figure}

MOF-5, also known as IRMOF-1, is a prototypical metal--organic framework, one of the first synthesized\cite{MOF5} and widely studied ever since. Its secondary building units consist of M$_4$O tetrahedra, with a central oxygen surrounded by 4 divalent cation (M\ex{2+}) sites forming a tetrahedron, and 1,4-benzenedicarboxylate linkers (abbreviated as ``bdc''). Each edge of the tetrahedron faces a carboxylate group from a linker, with each oxygen coordinating one of the two edge's cations, see Fig.~\ref{2clus2cell}(a). Linkers, connecting neighboring tetrahedra, are oriented along either of 3 axes orthogonal to each other, so that the MOF-5 structure has cubic symmetry (space group $Fm\bar{3}m$; see Fig.~\ref{2clus2cell}(b)).

The MOF-5 framework can \emph{a priori} sustain various types of divalent cations: Zn\ex{2+} is the cation present in the ``original'' MOF-5, but variants with other metals such as Be\ex{2+}, Mg\ex{2+}, Ca\ex{2+}, Cd\ex{2+}, \ldots have been considered theoretically\cite{Fuentes05, Yang2011} and synthesized experimentally.\cite{Hausdorf10, Brozek12} Conditions for polymetallicity in MOF-5, i.e. the coexistence of several cation types in the crystal structure, have been investigated in several recent experimental studies, both for fundamental aspects (regarding chemical and mechanical stability of MOFs) and motivated by possible applications e.g. in catalysis\cite{Brozek13} and adsorption.\cite{Botas10}

In this section we deal with MOF-5 structures where two types of divalent cations (Zn\ex{2+}, Cd\ex{2+}, Mg\ex{2+}, Ca\ex{2+}, Sr\ex{2+}, Ca\ex{2+}, and Be\ex{2+}) occupy the cationic sites. For various such pairs, we consider all bimetallic structures obtained from substitutions within the primitive cell of MOF-5 (see Section~\ref{sec:methods}). Our goal is to find out whether and which bimetallic patterns are energetically more favorable than the homometallic ones. We show how this depends not only on substitution patterns but also on intrinsic properties of the cations considered, such as ion size and electronic structure.

\subsection{Mixing with minimal strain effects: (Zn, Mg) substitution}

\begin{figure*}[t]
\begin{center}
\includegraphics[height=65mm]{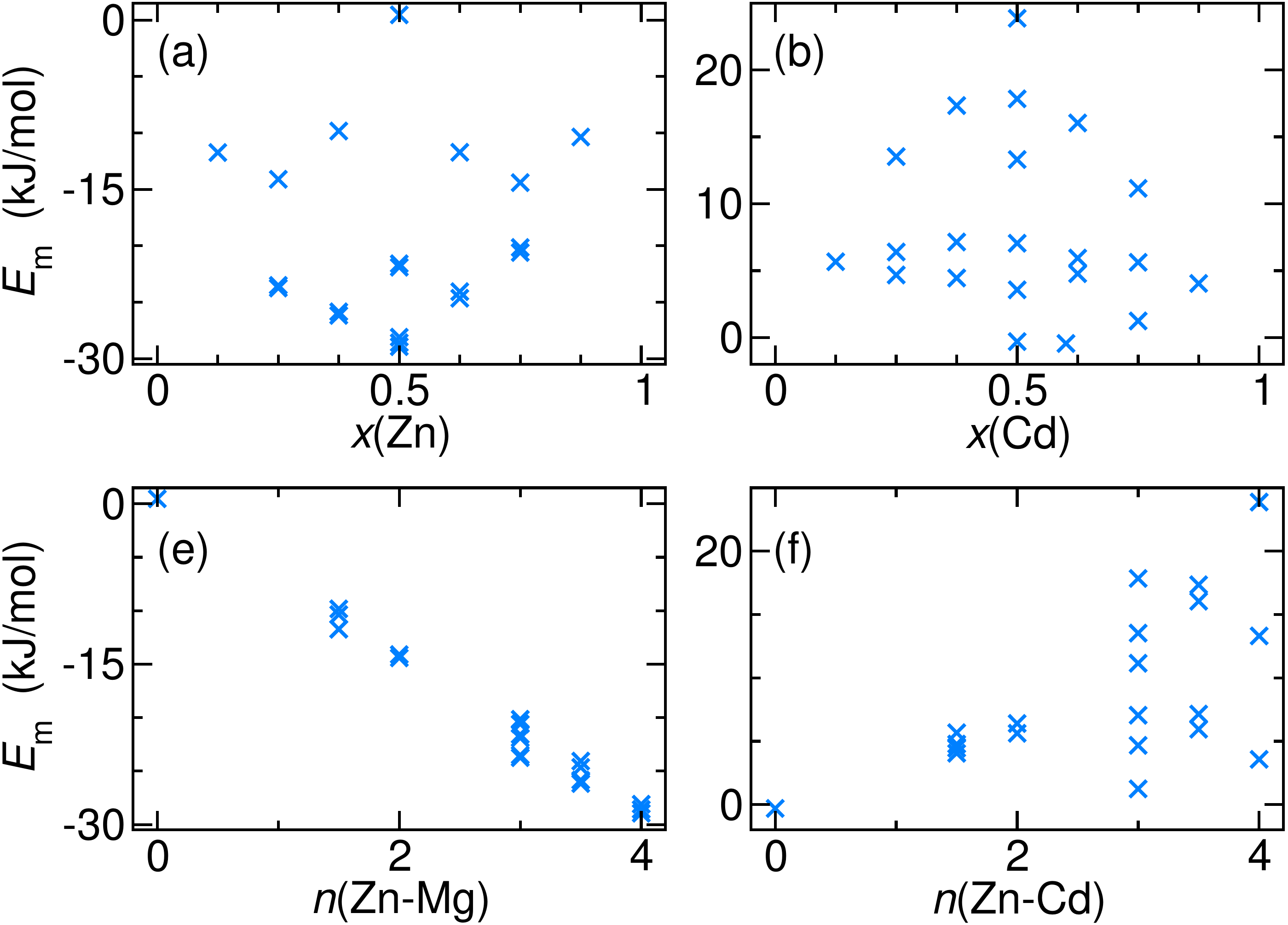}
\includegraphics[height=65mm]{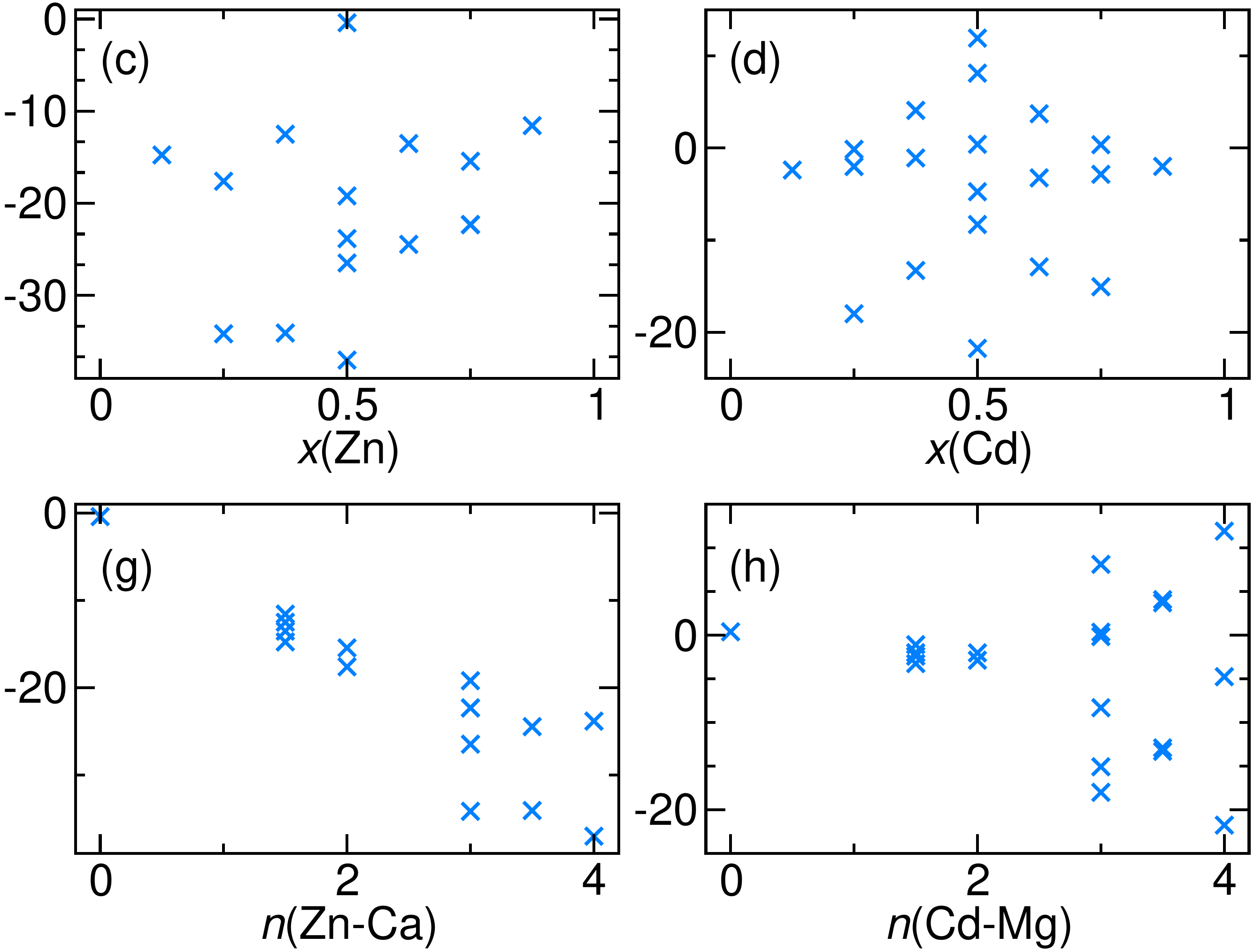}
\caption{\label{xnAB}Bimetallic MOF-5 structures: mixing energy $E_m$ (in kJ.mol$^{-1}$) versus (a--d): the fraction $x$(Zn) or $x$(Cd), of Zn or Cd respectively, at cationic sites; and (e--h) the number $n$(AB) of mixed tetrahedra edges (per cluster). Pairs of metal elements considered: (a,e): Zn and Mg; (b,f): Cd and Zn; (c,g): Zn and Ca; (d,h): Cd and Mg. }
\end{center}
\end{figure*}

The first situation we address is the simplest, namely that of mixing between two cations with similar size. We use here the example of Zn and Mg: their ionic radii are close (0.60~{\AA} and 0.57~{\AA} respectively\cite{Shannon1976}) and so are the lattice parameters and interatomic distances in the respective MOF-5 derivatives (e.g., 1.94~{\AA} and 1.96~{\AA} for the M--O distances; see Table~S1). For this pair of elements, the mixing energies are reported as a function of composition in Fig.~\ref{xnAB}(a). They are always negative, with values comprised between $-10$ and $-30$~kJ/mol (per primitive cell of 2 clusters), except in the case where both clusters are homometallic (there, $|E_m| < 1$kJ.mol$^{-1}$). The overall symmetry of the plot indicates that the mixing energies are nearly invariant upon interchanging the role of the elements. This is also seen upon metal exchange in isolated, formate-capped clusters (see Table~\ref{isoclus}).

\begin{figure*}[t]
\begin{center}
\includegraphics[height=60mm]{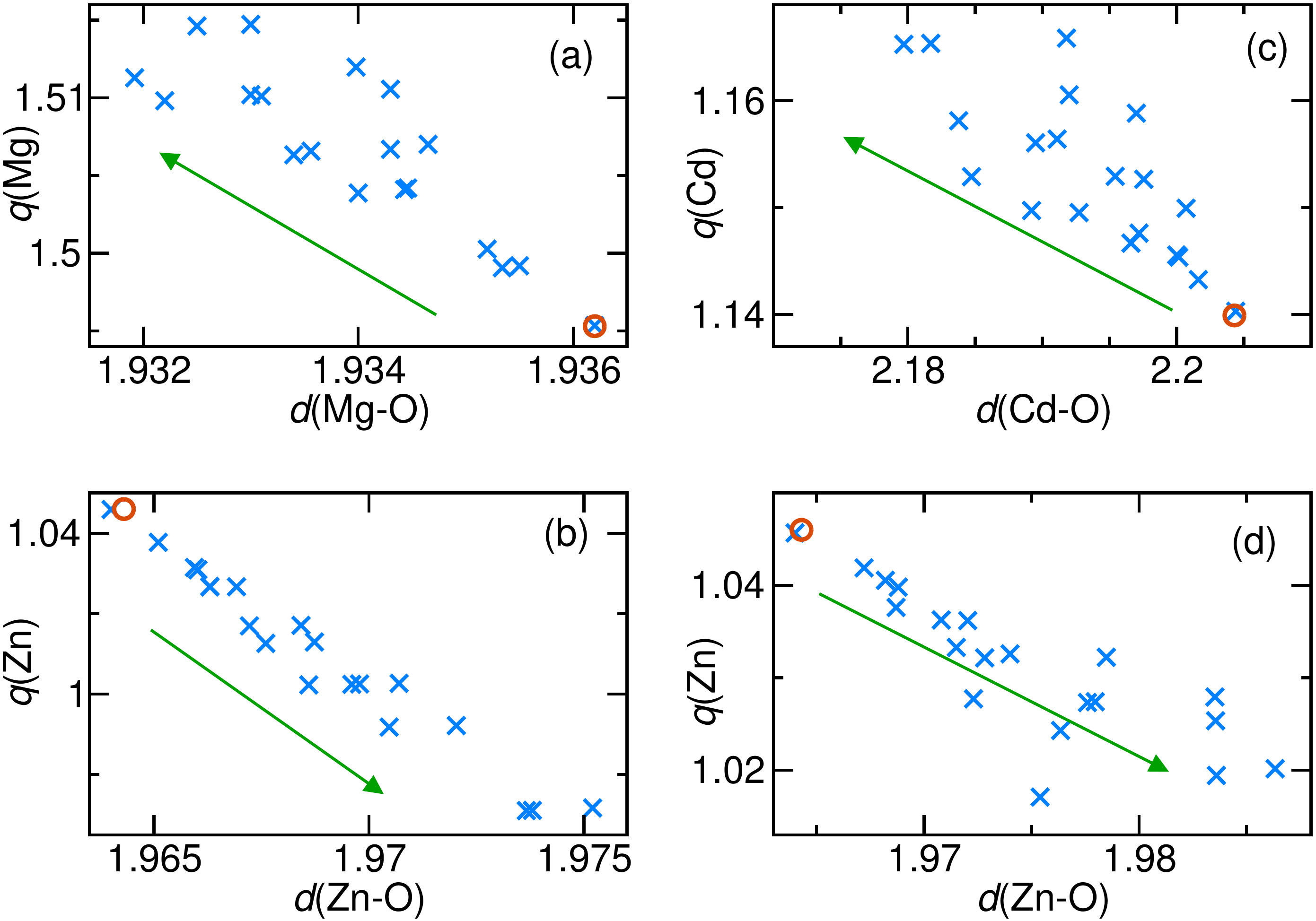}\hskip 3mm
\includegraphics[height=60mm]{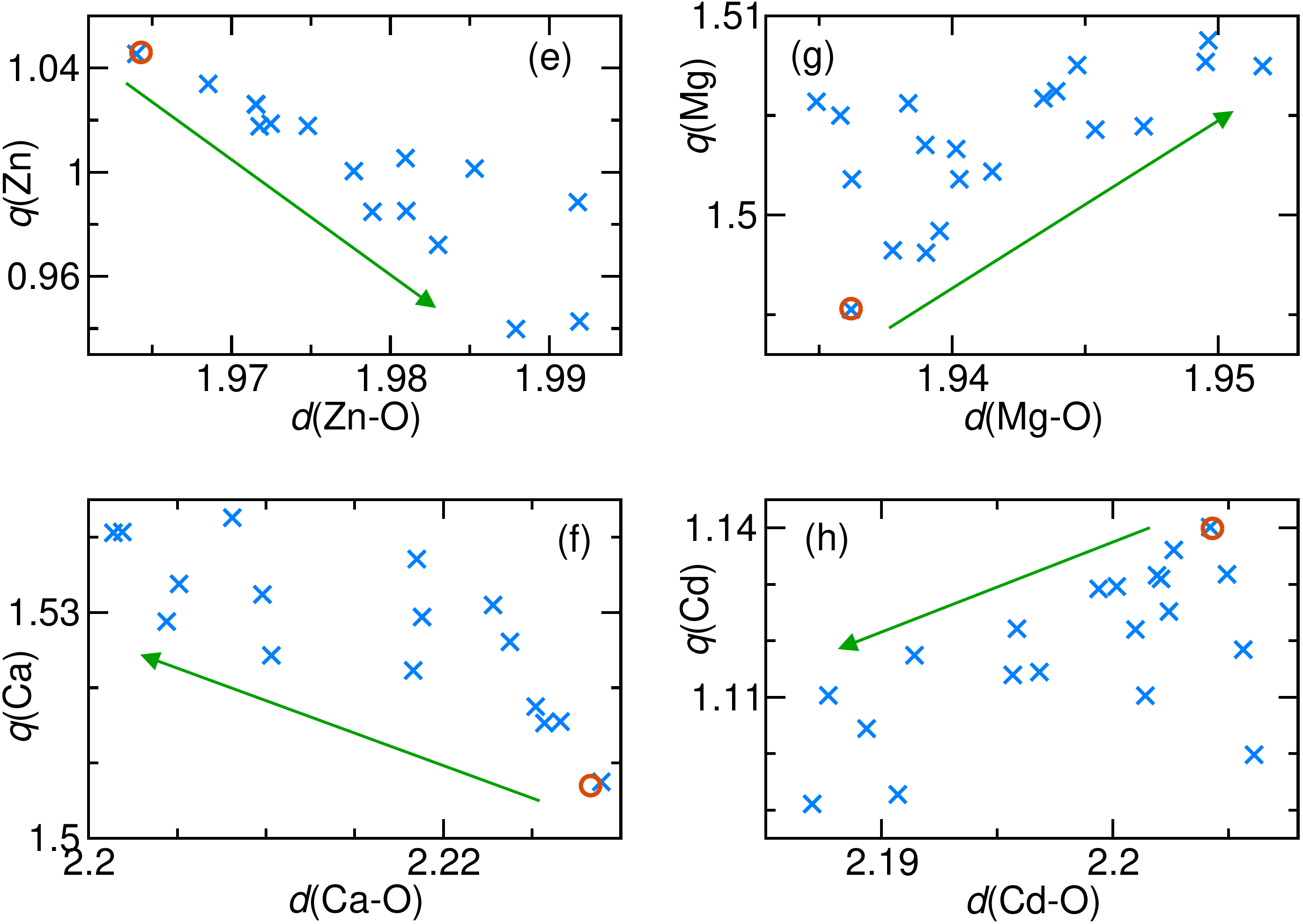}
\caption{\label{qdOc}(A,B) bimetallic MOF-5 systems: Evolution of net charges $q\e{M}$ (Mulliken atomic charges, in units of $e$) carried by the cations M = A, B versus distances $d\e{O--M}$ to the coordinating carboxylate oxygens (in \AA). Both quantities are spatially-averaged. Pairs of elements (A,B) coexisting in the strutcures: (a,b): Mg and Zn; (c,d): Cd and Zn; (e,f): Zn and Ca; (g,h): Mg and Cd. Points with red circles correspond to homometallic structures. The general trend upon substitution is highlighted by a green arrow.}
\end{center}
\end{figure*}

We show in Fig.~\ref{xnAB}(e) a plot of the mixing energy $E_m$ versus the number $n$(AB) (here A\,=\,Zn and B\,=\,Mg) of \emph{mixed tetrahedra edges}, i.e. the number of edges, in a tetrahedral metal cluster, that feature different cations at both ends. The clear linear correlation shows that the mixing energy is thus mainly determined by intra-cluster effects, namely of sum of pairwise interactions between neighboring cations. Furthermore, looking at metal-carboxylate coordination distances $d\e{Mg--O}$ and $d\e{Zn--O}$ in Fig.~\ref{qdOc}(a,b), one observes a surprising feature: upon mixing, the \emph{smaller} ion (Mg\ex{2+}) gets closer to its surrounding carboxylate oxygens, while the opposite occurs for Zn\ex{2+} --- though the magnitude of the effect is small, with changes of 1~pm at most. Further insight comes from correlating these distances with the charges on the respective cations: for both, the charge $q\e{M}$ decreases with increasing $d\e{M--O}$. This trend can be understood qualitatively within a classical picture of point charges.\footnote{Although net charges on oxygens may show variations upon mixing of the same order as those on cations, we checked (see Fig.~S4) that these variations depend mostly on the global composition. They are either much smaller than charge variations on cations, or seem uncorrelated with the energy.} This also means that the Zn/Mg charge difference, as that between the respective coordination distances, is increased upon mixing --- the corresponding value for pure structures being quite large already, 0.45 $e$ from Table~S1. This can be related to the ionization energies $E_{2e}$ of Zn and Mg \cite{Eion}, i.e. for an element X the energy cost of:
\begin{equation}
\text{X} \rightarrow \text{X}^{2+} +2\,e^-
\end{equation}
Indeed, the double ionization energy of Zn exceeds that of Mg by almost 3~eV. Thus, when both elements coexist in the same Zn$_n$Mg$_{4-n}$O SBU, one may expect a small electron transfer from Mg atoms (from which valence electrons can be more easily removed) to Zn atoms. Possible mechanisms for this charge transfer (direct exchange, superexchange involving the central O, or else) are not the purpose of this study; yet it seems to have an intrinsically chemical origin, and in any case it allows the system to gain energy from mixing, proportionally to the number of mixed tetrahedra edges.

\begin{table*}[t]
\renewcommand{\arraystretch}{1.4}
\setlength\tabcolsep{8pt}
\begin{tabular}{c|c|c|c|c|c|c|c|c|}
\multicolumn{2}{c}{} & \multicolumn{7}{c}{\textbf{cation A}} \\ \cline{3-9}
\multicolumn{1}{c}{} & & Be & Mg & Ca & Sr & Ba & Zn & Cd \\ \cline{2-9}
\multirow{7}{1.2ex}{\rotatebox{90}{\textbf{cation B}}}
& Be & --- & $-5.4$  & $-24.3$ & $-63.7$ & $-54.7$ & $-1.1$ & $-5.1$ \\ \cline{2-9}
& Mg & $-15.4$ & --- &  $-0.1$ & $-10.2$ & $-8.6$  & $-12.8$ & $-12.7$ \\ \cline{2-9}
& Ca & $-49.3$ & 0.05 & ---  & $-38.2$ & $-24.6$ & $-20.9$  & $-23.0$ \\ \cline{2-9}
& Sr & X     & $-17.4$ & $-33.0$ & --- & $-58.8$ & $-41.4$  & $-84.4$ \\ \cline{2-9}
& Ba & X     & 1.0  & $-18.5$  & $-17.8$ & --- & $-29.6$  & $-18.6$ \\ \cline{2-9}
& Zn & $-4.2$  & $-10.8$ & $-18.0$ & $-32.1$ & $-8.3$ & ---  & $-1.1$ \\ \cline{2-9}
& Cd & $-5.1$ & $-10.4$ & $-22.9$ & $-29.9$ & $-6.6$ & $-1.2$ & --- \\ \cline{2-9}
\end{tabular}
\caption{\label{isoclus}Mixing energies $E_m$ (in kJ.mol$^{-1}$) for A\e{3}B clusters of one MOF-5 tetrahedral metal center capped with formate linkers. Symbol X (for the Sr$_3$Be and Ba$_3$Be cases) signals DFT calculations that yielded physically unrealistic structures where the cluster integrity is not retained.}
\end{table*}

\subsection{(Cd, Zn) bimetallic systems: effects of size mismatch}

A radically different situation is found in (Cd, Zn) mixed frameworks, where both cations are of similar chemical nature, featuring $d^{10}$ electronic configurations and similar atomic charges in the homometallic MOF-5 framework (see Table~S1). Yet they clearly differ in their size, and from this we may expect an important effect of the framework deformation to appear in the mixing energetics. As in the case of the (Zn, Mg) system, the plot of mixing energies represented in Fig.~\ref{xnAB}(b) is almost symmetric. Yet in this case the mixing energies are generally positive, showing that heterometallic (Cd, Zn) MOF-5 are energetically unstable compared to the separate homometallic phases, i.e. mixing these two cations costs energy.

If we plot the mixing energy $E_m$ as a function of the number $n$(AB) of mixed tetrahedra edges, it doesn't have the linear behavior previously observed in the case of (Zn, Mg). This reflects the importance of lattice strains, induced by the difference in cation sizes at a positive energy cost; in each structure, strains depend on positions of all cations, and not only on the number of neighboring cations of different type. This is also reflected in the mixing energies for the isolated metal clusters (Table~\ref{isoclus}): while the mixing energy for formate-capped Zn$_3$MgO and ZnMg$_3$O clusters were both clearly negative ($-12.8$ and $-10.8$~kJ/mol respectively), those of the Zn$_3$CdO and ZnCd$_3$O clusters are much smaller ($-1.1$ and $-1.2$~kJ/mol) due to larger deformation of tetrahedra.

We thus wished to identify some key descriptors of the topologies of heterometallic structures (i.e. quantities depending on the Cd or Zn occupation of each cationic sites, but independent of the mixing-induced relaxations) that may explain better the observed $E_m$ values. For that, let us imagine linkers as rigid bodies --- an approximation actually reasonable when considering intra-linker bond distances and angles in relaxed structures.\footnote{Mean standard deviations of linker bond lengths, measured on all relaxed structures, are $\lesssim 0.5$~pm; C--C=C bond angles deviate from 120 degrees by at most 2 degrees while C--O--C and intra-phenyl C--C--C bond angles show even less variation.} A linker's position and orientation will relax more or less efficiently depending on its coordinating environment, i.e. on which cation is coordinated by each of the 4 oxygens. For instance, if the linker is coordinated by 2 cations of each type, one at each COO$^-$ group in \emph{trans} position, it may adapt to its environment by a small rotation around the benzene's $C_6$ axis. In contrary, if the linker coordinates 2 cations of each type in \emph{cis} configuration, such a rotation doesn't help, so coordination and other bonds are expected to be more strongly distorted, at a higher energy cost.

\begin{figure}[t]
\begin{center}
\includegraphics[width=79mm]{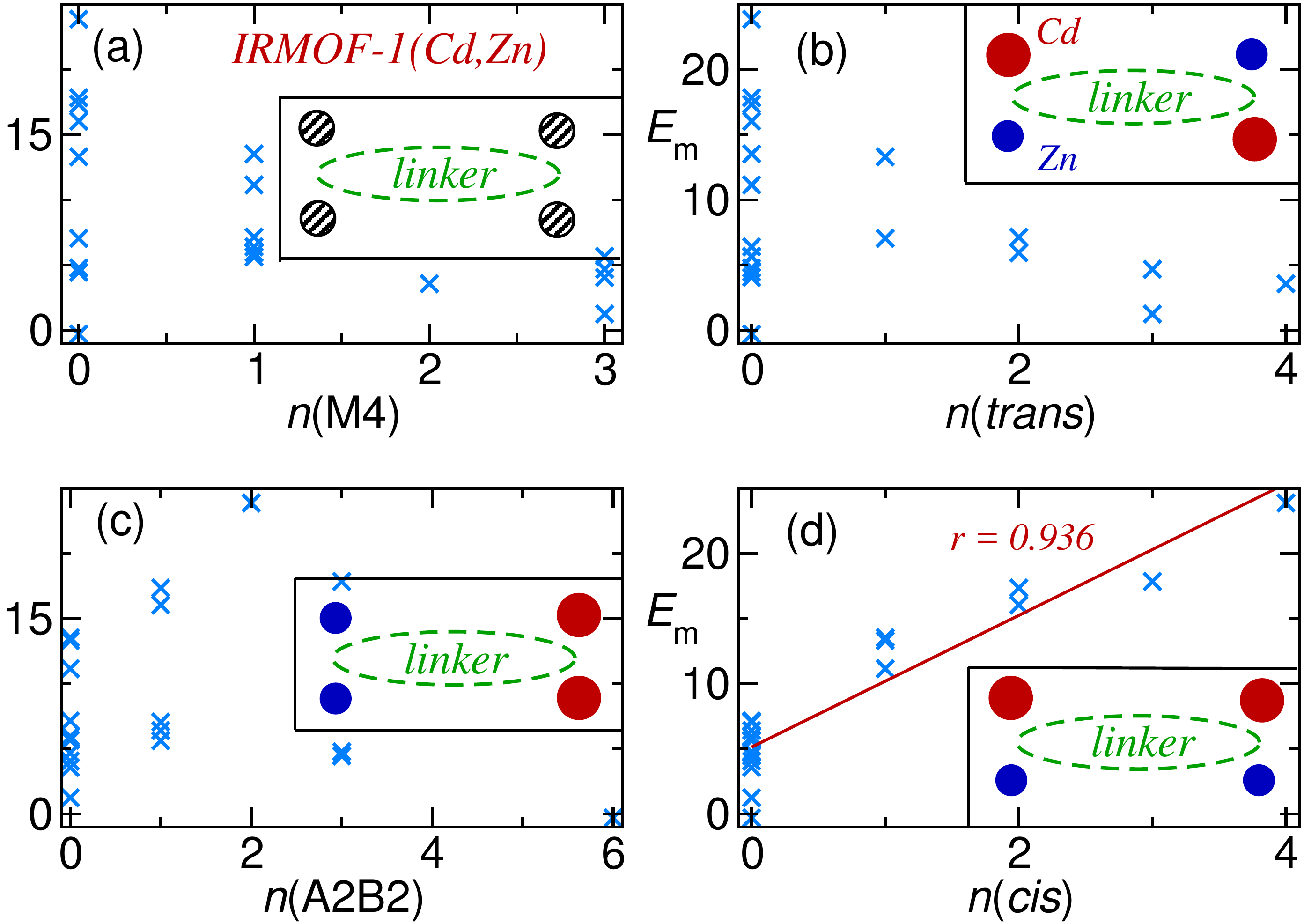}\\[2mm]
\includegraphics[width=79mm]{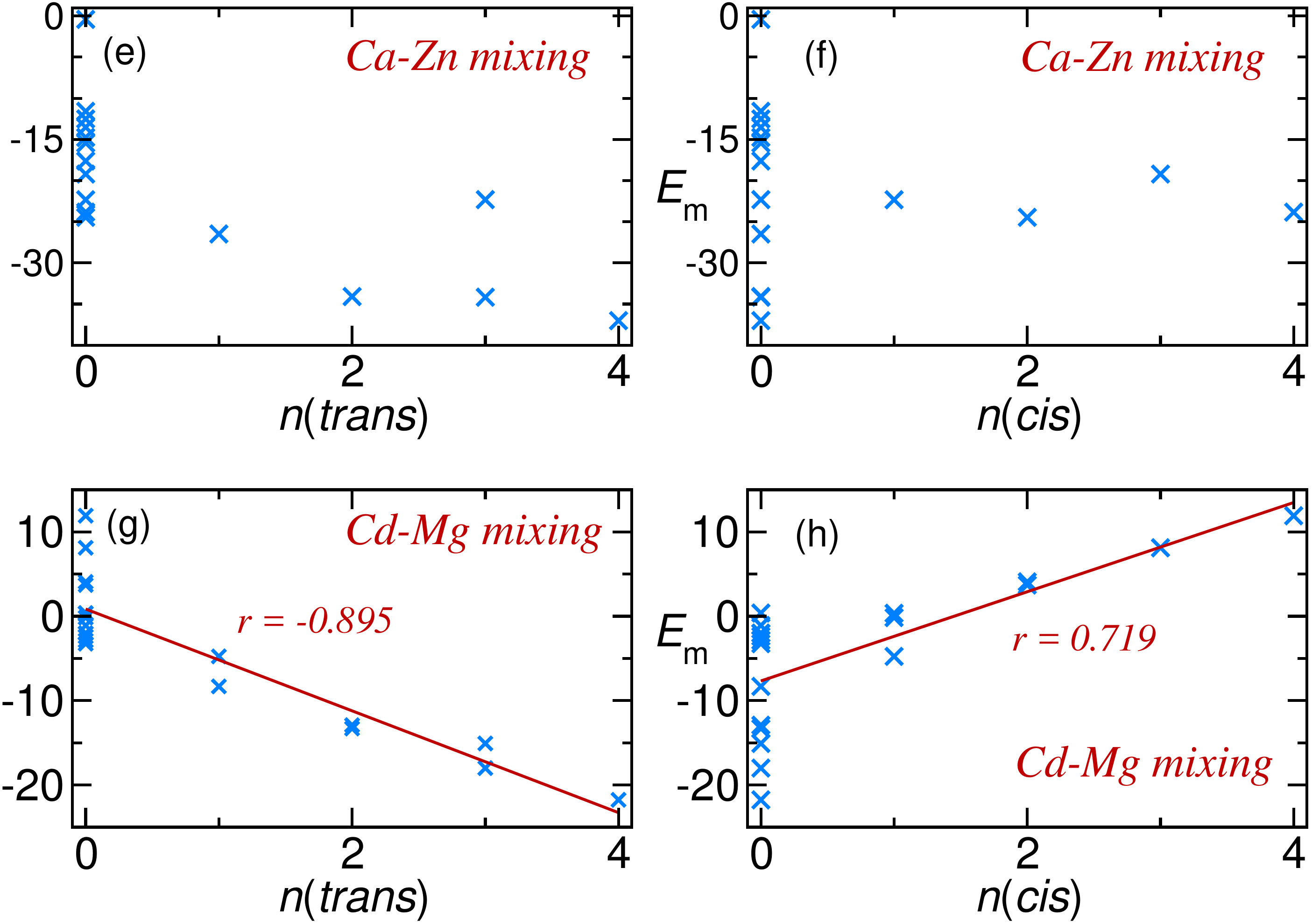}\\
\caption{\label{linkers} Bimetallic MOF-5: mixing energy $E_m$ (in kJ.mol$^{-1}$) versus number of linkers in various configurations. (a-d): for (Cd, Zn) mixed structures; (e,f): (Ca, Zn); and (g-h): (Cd, Mg) mixed structures. Abscisses $n$(M4), $n$(A2-B2), $n$(\emph{trans}) and $n$(\emph{cis}) refer to the configurations depicted in insets of (a-d).}
\end{center}
\end{figure}

Indeed, when plotting $E_m$ versus several types of linker descriptors, in Fig.~\ref{linkers}(a-d), one sees a quite clear correlation with the numbers $n$(\emph{cis}) and $n$(\emph{trans}) of linkers in \emph{cis} and \emph{trans} configuration respectively -- whereas the energy doesn't seem to depend on other linker descriptors, such as the number $n$(M4) of linkers coordinating 4 cations of the same type. To go further, we performed a multi-variable analysis, assuming a law of the form $E_m\ex{$(j)$,pred} = a_1 n(\text{AB})^{(j)} + a_2 n(\text{\emph{cis}})^{(j)} + a_3 n(\text{\emph{trans}})^{(j)}$ (see Supporting Information for details). This provides optimal coefficients $a_i$ defining a predicted $y=E_m\ex{$(j)$,pred}$, such that a linear regression of $E_m^{(j)}$ versus $E_m\ex{$(j)$,pred}$ gives a very good correlation coefficient ($r > 0.99$). The predicted mixing energy is, as expected, increased by heterometallic edges ($a_1=3.2$), and by \emph{cis} linkers ($a_2=2.8$) but (to a smaller extent) decreased by \emph{trans} linkers ($a_3=-2.2$).

In contrast to the (Mg, Zn) case, coordination distances $d\e{Zn--O}$ of the smaller cation (Zn) are larger in mixed structures than the value $d^0\e{Zn--O}$ in pure MOF-5(Zn), while $d\e{Cd--O}$ is decreased upon mixing [see Fig.~\ref{qdOc}(c,d)]. This indicates a repartition of mixing-induced strains between these two types of bonds, in order to minimize the strain-induced energy cost. Note that other bond/angle degrees of freedom can also relax in the mixing process, such as for example cation--cation distances in a cluster (consistently with the rather large, positive $a_1$ value).

Charges on both cations show a similar behavior as in the (Mg, Zn) case, namely a rough decrease with increasing coordination distance, as can be seen on Fig.~\ref{qdOc}(c,d). Again, the (Cd, Zn) charge disparity is increased upon mixing. Yet here, considering the small difference ($<1$~eV) between $E_{2e}$ values of both ions, the observed charge transfer might have another origin. It actually seems to result, at least partly, from the variations in coordination distances (see previous paragraph), to which the charge degrees of freedom adapt.  Assuming that the system gains energy from charge transfer, in (Cd, Zn) bimetallic systems its amplitude is too small for the resulting energy gain to balance the costs of mixing-induced strains, hence explaining that these heterometallic structures are energetically unfavorable ($E_m>0$).

\subsection{Charge transfer versus cation size mismatch}

Having identified on the cases of (Mg, Zn) and (Cd, Zn) pairs two main mechanisms impacting on the mixing of divalent cations in MOF-5, namely lattice strains induced by the difference in cation sizes and charge transfer with intrinsically chemical origin, we then address more generic situations, when both these effects come into play. For this we now turn to the two cases of (Ca, Zn) and (Cd, Mg) heterometallic MOF-5 structures. In both cases, the first ion (A = Cd or Ca) is clearly larger than the second (B = Zn or Mg). Thus, in analogy to the (Cd, Zn) case, mixing-induced strains are expected to lead to a reduction of the $d\e{A--O}$ coordination distances, and to increase the charges $q\e{A}$. Yet these two situations differ when considering the double ionization energies of elements involved. For (Ca, Zn) systems, since E$_{2e}$(Zn)$\gg$E$_{2e}$(Ca), one also expects an increase in $q\e{Ca}$ and decrease in $q\e{Zn}$; while for (Cd, Mg) systems, E$_{2e}$(Cd)>E$_{2e}$(Mg) so in absence of mixing-induced strains one would expect a decrease in $q\e{Cd}$ and increase in $q\e{Mg}$.

In the case of (Ca, Zn), mixing energies seen on Fig.~\ref{xnAB}(c) are found always negative (for the 16 out of 20 mixed structures that relaxed successfully) and not fully invariant upon Zn$\leftrightarrow$Ca interchange. $E_m$ again depends mainly on the number of mixed tetrahedra edges, but with somehow an influence of the role of linker descriptors like $n$(\emph{trans}) [see Fig.~\ref{linkers}(e)]. A multi-variable analysis, similar as that done above, confirms this with a coefficient $a_1=-8.9$ [associated to $n$(AB)] while other $|a_i|$ values are at least 3 times smaller. Coordination distances and net charges, shown in Fig.~\ref{qdOc}(e,f), behave similarly as in the (Cd, Zn) case, with e.g. a charge decrease for the (smaller and harder to ionize) Zn ion, roughly proportional to an increase in $d\e{Zn--O}$. It thus appears that, when both size mismatch and charge transfer ``push'' in the same direction, they have a cooperative effect on the relaxation of atomic positions and charges. When, as in the present case, the charge transfer effect is strong enough, the resulting energy gain overcomes strain-induced energy costs, and allows for generally negative mixing energies and a solid-solution behavior.

In contrast, in the (Cd, Mg) system these two effects compete. In consequence, a more contrasted situation is observed in Fig.~\ref{xnAB}(d,h), with $E_m$ taking both positive and negative values depending on the composition and configuration. While its dependence on the number of mixed edges is unclear (Fig.~\ref{xnAB}(h)), the mixing energy clearly tends to increase with $n$(\emph{cis}) and to decrease with $n$(\emph{trans}), as was also seen e.g. for (Cd, Zn). Again, a multi-variable analysis confirms this trend: $E_m$ behaves almost linearly in function of an optimal variable $y$ built on $n$(\emph{cis}) and $n$(\emph{trans}) only, with a correlation coefficient $r=0.992$. The net charges on both Cd and Mg follows their ionization potentials (decrease in $q\e{Cd}$ and increase in $q\e{Mg}$), but with amplitudes typically smaller than in the (Mg, Zn) or (Ca, Zn) systems. Thus the energy gain from charge transfer is smaller, but may be sufficient to have $E_m<0$ if enough linkers have environments allowing for efficient relaxations (e.g. many linkers in \emph{trans} configuration and few in \emph{cis} configuration). Aside from the structure-dependent sign of $E_m$, a further indication of the competition between both effects, frustration of charge and strain degrees of freedom, is that cation charges $q$(M) do not decrease with increasing coordination distance (see Fig.~\ref{qdOc}(g,h)) in contrary to the previous cases.

\subsection{Other bimetallic MOF-5 systems}

We presented in the previous sections some examples of bimetallic MOF-5 systems with representative behavior. We performed further simulations on additional bimetallic systems in order to confirm the conclusions reached. In particular, results of calculations on periodic systems for four other choices of cationic element pairs are shown in the Supporting Information. They all exhibit similarity with the archetypical cases exposed above. (Cd, Ca) bimetallic systems, where both cations are of nearly same size, show a solid-solution behavior similar to the (Zn, Mg) case. (Zn, Sr) systems show similarities with the (Zn, Ca) case, with even lower mixing energies and larger displacements due the even larger mutual differences in both charges and sizes of cations. In (Ca, Mg) and (Be, Mg) systems, which are characterized by large differences between cation sizes, mixing energies are very sensitive to linker environments. In particular, in the (Ca, Mg) case where the mutual charge difference is negligible, they correlate well with changes in coordination
distances.

In addition to these periodic calculations, we performed systematic calculations on isolated bimetallic A$_3$BO clusters, comprising 3 cations of one element and one of another element, and with 1,4-benzenedicarboxylate linkers replaced by formates to cap the cluster. We performed these calculations for all couples where A and B are either: Be, Mg, Ca, Sr, Ba, Zn, or Cd --- hence a total of 42 different clusters. 40 of these lead to a stable structure, retaining the integrity of the inorganic cluster. We report in Table~\ref{isoclus} their mixing energies. We can see that cation mixing is generally favorable, with a negative mixing energy $E_m$. It is however less favorable when both cations bear similar partial charges, such as the cases of (Ca, Mg) and (Zn, Be), confirming that the charge imbalance, taken alone, drives an electronic reconstruction thanks to which the energy is lowered upon mixing. For most pairs of elements (A, B), mixing energies $E_m$(A$_3$B) and $E_m$(B$_3$A) are in the same range, but with small relative differences --- apart from systems with cations of very different size or nature, such as (Ba, Sr) or (Ba, Zn). We can thus see that while a wide variety of behavior can be expected in heterometallic MOFs, in the absence of strain the mixing of metal cations is in a vast majority of cases energetically favorable.

\section{Bimetallicity in UiO-66\label{sec:UiO}}

The UiO-66 structure (see Figure~\ref{2clus2cell}) is formed with M$_6$O$_4$(OH)$_4$ octahedra (later abbreviated M$_6$), where M$^{4+}$ is a tetravalent cation (M = Ti, Zr, Hf or Ce), connected to each other here again by \emph{bdc} linkers. Materials of the UiO-66 family are, like IRMOFs, an interesting playground for studying the consequences of cation substitutions and possibility of bimetallicity. Partial Zr$\,\rightarrow\,$Ti or Zr$\,\rightarrow\,$Hf post-synthetic exchange (PSE) have been reported in the literature \cite{Kim12}, with more efficiency in the first case. In both cases the bimetallic samples retained the UiO-66 structure, however no information could be obtained on the spatial distribution of cations. It is not known whether intra-cluster mixing occurs, whether specific substitution patterns are favored, in a word whether some order exists within the heterometallic samples. Post-synthetic replacement of Zr by Ce has also been achieved in UiO-66, \cite{Nouar15} but this study revealed a change in oxidation state during the process, as well as a possible influence of ligand vacancies. Regarding possible applications, partial post-synthetic exchange in UiO-66 was shown to improve its efficiency in catalysis,\cite{Nouar15} as well as in CO$_2$ adsorption.\cite{Lau2013}

\begin{figure}[t]
\begin{center}
\includegraphics[width=7cm]{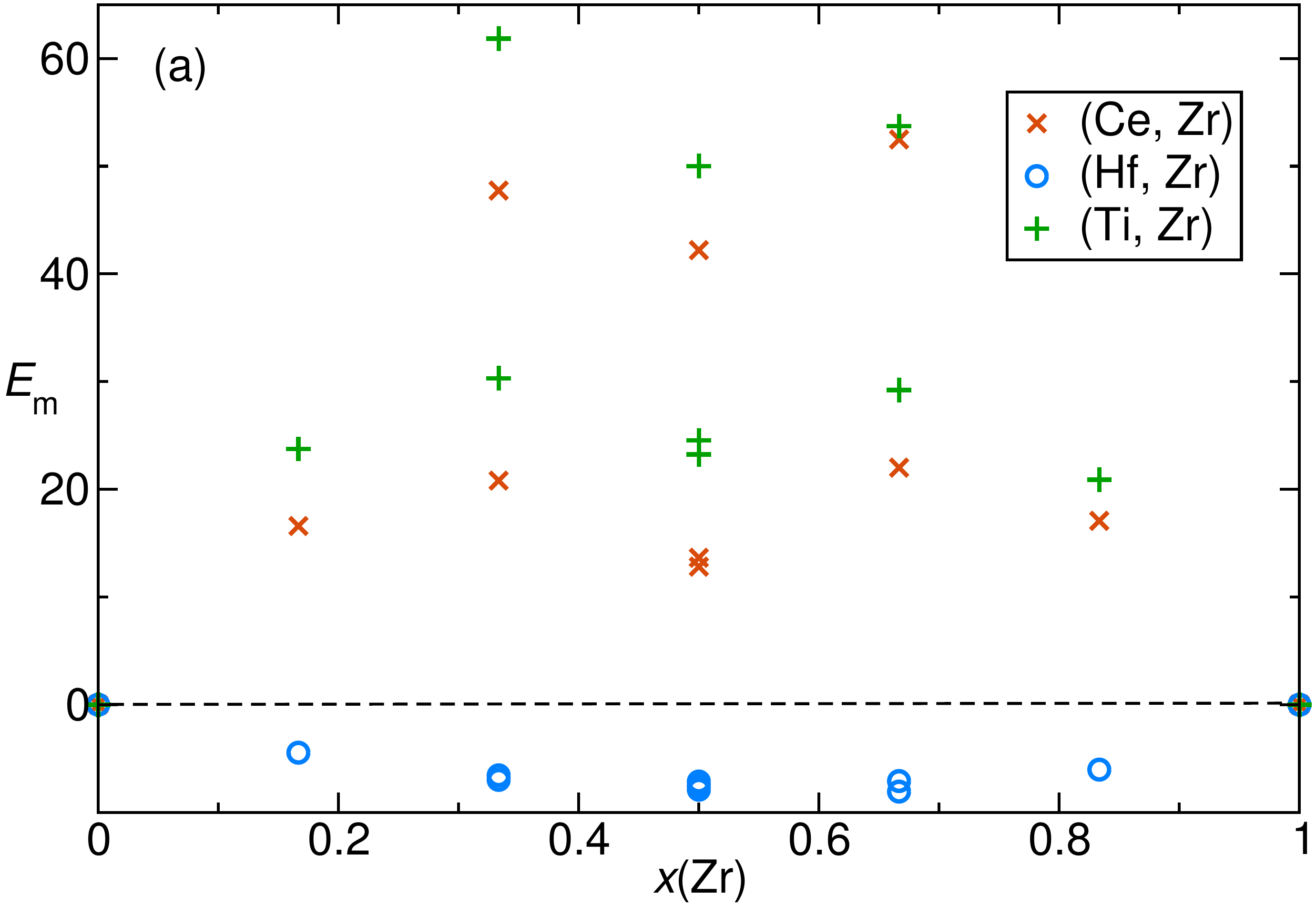}\\[2mm]
\includegraphics[width=7cm]{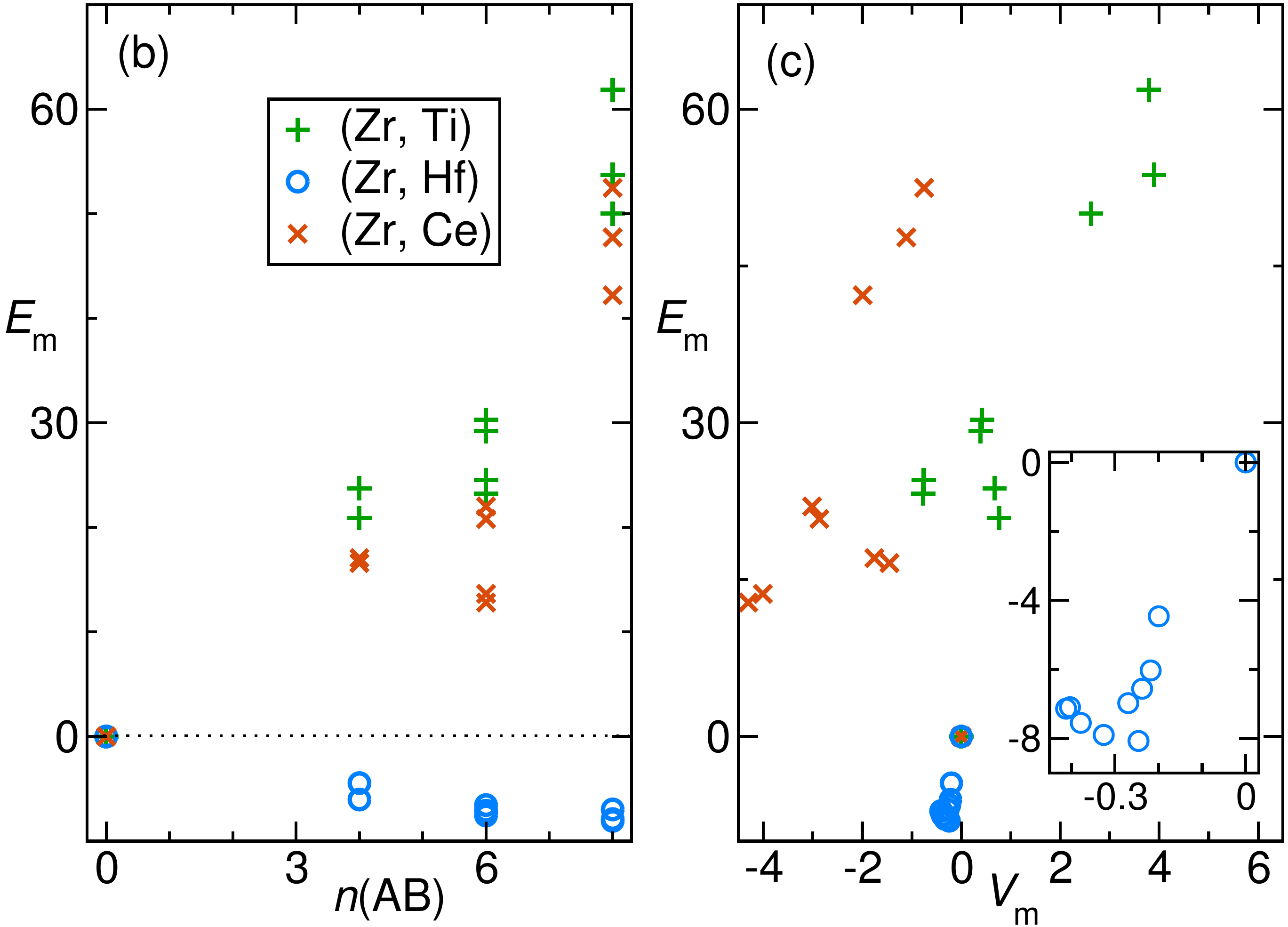}
\caption{\label{UiO_intra}Bimetallic UiO-66 structures (substitutions within a primitive cell), with coexistence of Zr and either Ti, Hf or Ce. Top: mixing energy $E_m$ versus Zr occupancy on cationic sites. Bottom: $E_m$ versus (b) number of bimetallic octahedra edges $n$(AB) (per octahedron) and (c) mixing-induced variation of the primitive cell volume (in \AA$^3$).}
\end{center}
\end{figure}

\subsection{Substitutions in a primitive cell\label{sec:UiOprim}}

The standard UiO-66 structure has cubic symmetry and a $F\bar{4}3m$ space group, with a primitive unit cell containing a single Zr$_6$ octahedron while the conventional cell contains four. If considering, as earlier for MOF-5, cation substitutions within a primitive cell, one obtains $2^6 = 64$ possible substituted structures; but taking into account point group symmetries reduces the number of symmetry-inequivalent bimetallic structures to only 9 (excluding homometallic cases).

First we discuss bimetallic structures built from substitutions within a primitive cell of UiO-66. For three pairs of metallic elements: (Zr, Ti), (Zr, Ce) and (Zr, Hf), all such structures have been relaxed, and their mixing energies are depicted in Fig.~\ref{UiO_intra}. They are always negative for (Zr, Hf) systems, which is consistent with our previous observations in MOF-5, given the very close sizes of Zr and Hf. The roughly linear dependence of $E_m$ on the number of bimetallic octahedra edges, shown for (Zr, Hf) systems in Fig.~\ref{UiO_intra}(b), indicates a small mixing-induced energy gain, possibly from charge transfer. In contrast, mixing energies are positive for the other element pairs, (Zr, Ti) and (Zr, Ce), where mixing is expected to occur at a substantial deformation cost (see Table~S2). Given the quadruple ionization energies of the elements involved, charge transfer effects are also expected to occur and even be more important than for (Zr, Hf), but not sufficient to counterbalance deformation costs. A further indication for the role of these energy costs comes from a comparison with mixing energies computed on isolated clusters (see Table~S4): while in the latter case mixing energies are much lower for (Zr, Ce) clusters than for (Zr, Ti) clusters, the mutual difference is much reduced in periodic systems, presumably by lattice effects.

The mixing-induced volume variation, $V_m$, can be defined analogously to the mixing energy. Both quantities are plotted against each other in Fig.~\ref{UiO_intra}(c), for the systems considered above. They correlate rather well in the (Zr, Hf) case, with moderate mixing-induced reduction in cell volume nearly proportional to the mixing-induced energy gain. This correlation is less obvious in the (Zr, Ti) and (Zr, Ce) cases, which also differ qualitatively from each other: in the former the volume increases upon mixing in most structures, while in the latter it always decreases. This difference may have the same origin as the relatively lower mixing energies for (Zr, Ce) than for (Zr,Ti) systems, and indicate more efficient attractive interactions between neighboring cations in the former case.

Finally, linker configurations are expected to correlate with the mixing energies in bimetallic UIO-66 structures. Yet due to the small size of the cell considered (i.e. small number of mixed structures, with linker environments much constrained by periodicity) we could not draw precise conclusions on the role of linker configurations from these data alone -- this led us to consider a larger set of bimetallic structures, as described hereafter.

\subsection{Substitutions in a conventional cell\label{sec:conv}}

In order to better identify the key factors impacting the energy of bimetallic UiO-66 structures, and in particular the role of linker configurations, we focus here on the case of the bimetallic (Zr, Ce) UiO-66 and consider structures where substitutions were carried out in the conventional cell of UiO-66, containing four M$_6$ clusters. This allows for a much larger number of possible configurations (about $2^{24}/96 \simeq 175,000$ symmetry-independent structures). Among those, we focused on a subclass of 30 configurations, chosen to fulfill the following two criteria:
\begin{itemize}
\item all clusters, for fixed cluster composition, minimize the number of bimetallic octahedra edges; the motivation for this criterion is to focus on other factors influencing the energy, as well as on low- rather than high-energy structures;
\item retaining by at least one (and in most cases several) point group symmetry, in order to minimize computation time.
\end{itemize}

\begin{figure}[t]
\begin{center}
\includegraphics[width=7.6cm]{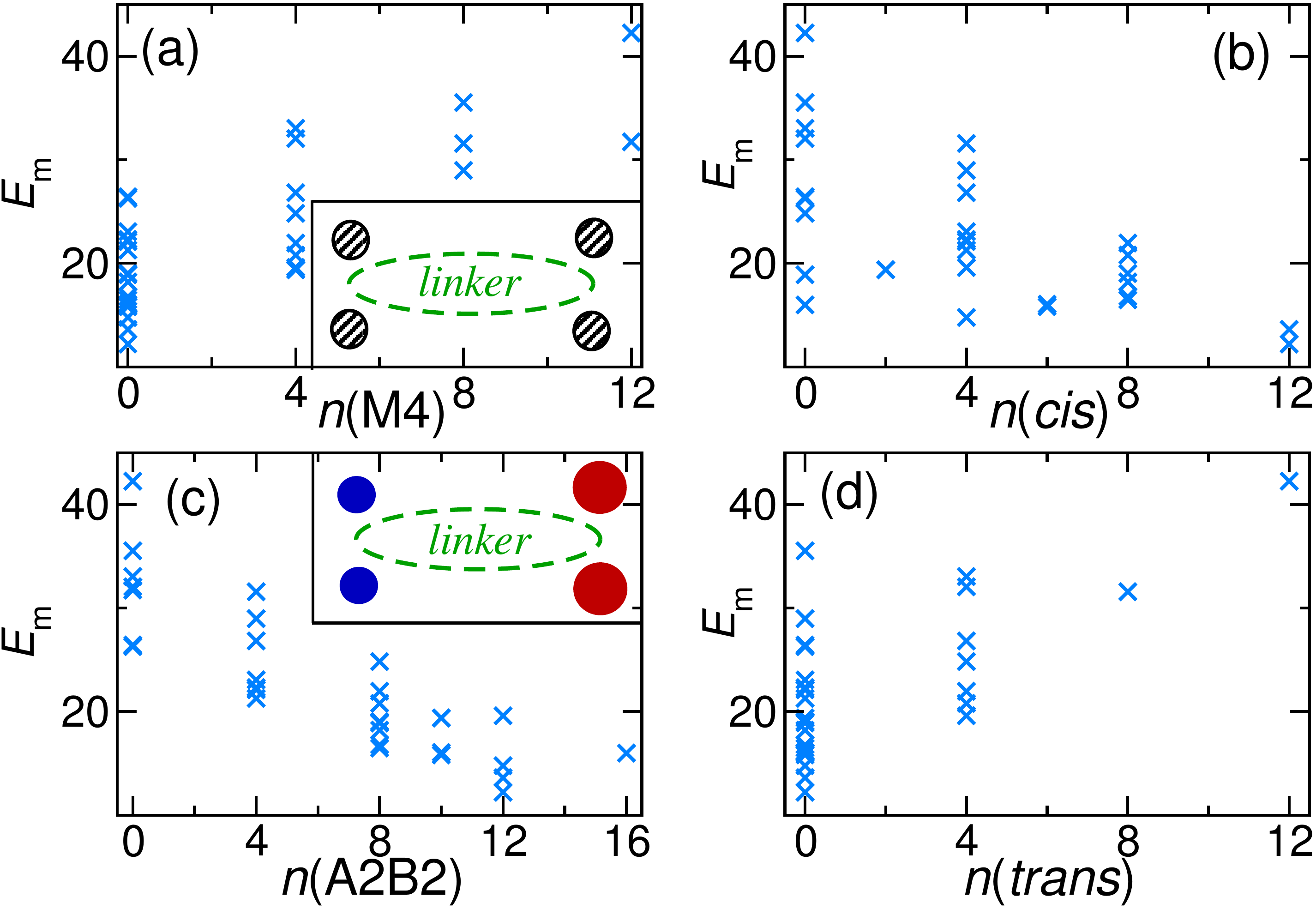}
\\[2mm]
\includegraphics[width=7.6cm]{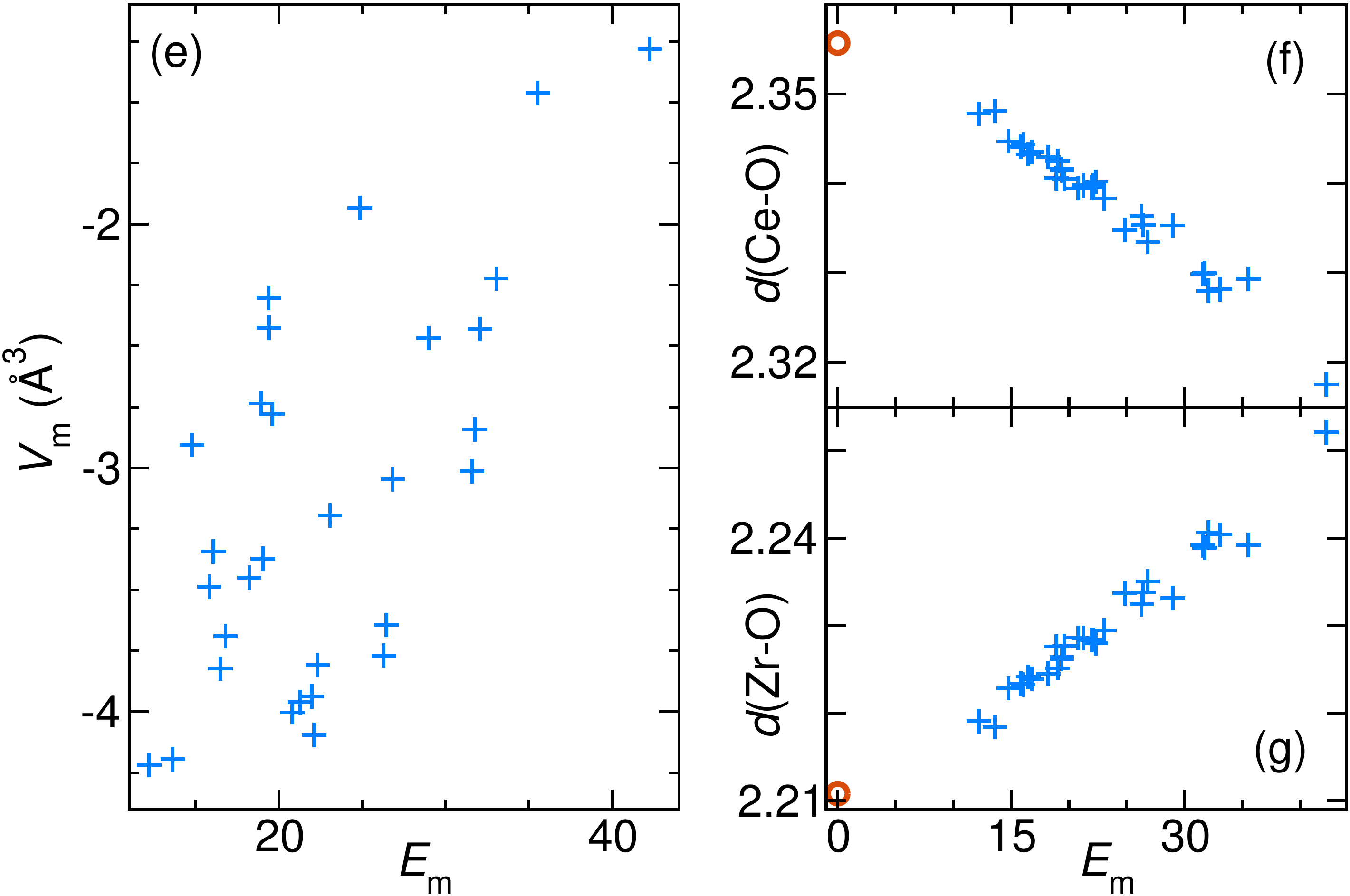}
\caption{\label{CeZr_conv} UiO-66(Ce,Zr) structures (substitutions within a conventional cell). Right: Mixing energy $E_m$ (in kJ.mol$^{-1}$, per cluster) versus numbers of linkers of specific types (as indicated in Fig.~\ref{linkers}) --- all quantities are per primitive cell. Left: Plots of $E_m$ (abscisses) versus: (e) mixing-induced volume variation $V_m$ (in A$^3$, per primitive cell); and (f,g): average coordination distances between either Ce (f) or Zr (g) and the carboxylate oxygens coordinating them (in \AA).}
\end{center}
\end{figure}

For all of the selected structures, we found positive mixing energies (Fig.~\ref{CeZr_conv}) in the range of 10--40~kJ.mol$^{-1}$ (per primitive cell), similar to those observed when substitutions were considered in a primitive cell. Yet here, the larger data set allowed us to identify clear trends concerning the impact of individual variables (number of bimetallic edges, linker configurations). Fig.~\ref{CeZr_conv}(a-d) shows a clear correlation between the energy and the numbers of 2 types of linkers: $E_m$ tends to be higher with more linkers coordinating 4 atoms of the same type [descriptor $n$(M4), Fig.~\ref{CeZr_conv}(a)], and lower with linkers coordinating 2 Ce on one carboxylate group and 2 Zr on the other [descriptor $n$(A2B2), Fig.~\ref{CeZr_conv}(c)]. It seems to correlate with the numbers of \emph{trans} and \emph{cis} linkers as well, yet less clearly -- same goes for the correlation between $E_m$ and the number $n$(AB) of bimetallic octahedra edges (see Fig.~S5).  A multi-variate analysis confirmed that the energy is more directly influenced by linker descriptors $n$(M4) and $n$(A2B2), rather than by other linker- or cluster-descriptors.

In other words, in bimetallic UiO-66 framework, effects of size mismatch --- when the two cations have significantly different size --- make the linker configurations energetically inequivalent, as in MOF-5, but in a different way. This is due to the higher cluster coordination (12 ligands around a cluster, instead of 6 in MOF-5). Indeed, linker rotations, which stabilize \textit{trans} linkers in MOF-5, are more difficult in UiO-66: such a rotation would bring a COO$^-$ group of the rotated ligand too close to a COO$^-$ group of a neighboring ligand, and thus involve significant additional linker--linker repulsions. Instead, the optimal linker configuration is the A$_2$B$_2$ configuration, where each COO$^-$ group coordinates one type of cations: the corresponding relaxation process, to adapt the cation size mismatch, is a small-amplitude translation along the linker axis. This does not bring the translated linker too close to another linker, and appears thus as the most efficient relaxation mechanism, compared to other linker configurations.

We note that bimetallic (Zr, Hf) systems behave quite differently in this respect (see Fig.~\ref{UiO_intra} and Fig.~S6), with the (negative) mixing energy nearly proportional to the number of bimetallic octahedra edges. There, even in a higly-coordinated framework, the quasi-absence of mixing-induced strains leaves intra-cluster interactions as the key mechanism dominating mixing energies.

For further insight, we also analyzed some variables quantifying the mixing-induced lattice distortions: relative volume variations, quantified by the mixing volume $V_m$, and coordination distances $d\e{Ce--O}$ and $d\e{Zr--O}$ (here again to carboxylate oxygens) and correlate them with the mixing energy. Fig.~\ref{CeZr_conv}(e) indicates that the mixing-induced volume variation is always negative (as seen in subsection~\ref{sec:UiOprim} for (Zr, Ce) systems), and the volume reduction tends to be more important for lower-energy structures. As for coordination distances, shown in Fig.~\ref{CeZr_conv}(f,g), they behave similarly as those of e.g. (Cd, Zn) MOF-5 structures: those of the smaller (Zr) ion increase upon mixing while $d\e{Ce--O}$ are decreased. Remarkably, we observe a clear linear correlation between those distances and the mixing energy. These features confirm the importance of lattice distortions in determining the mixing energies and relative stabilities of bimetallic structures, and the interplay between these distortions and linker environments.

\section{Conclusions and perspectives}

In this study, we have proposed and used a DFT-based methodology for a systematic study of heterometallic Metal-Organic Frameworks, exemplified on the frameworks MOF-5 and UiO-66. Based on criteria of energetic stability, we could determine, among the cation pairs considered, which ones can lead to stable bimetallic MOF phases, and adressed the spatial distribution of metals in such phases.

These results give a coherent picture of mixed-metal MOFs, with two essential aspects dominating their energetics. First, the coexistence of distinct types of cations inside a SBU (cluster) leads to a charge transfer between them, depending on their intrinsic chemical properties, and allows the system to gain energy upon mixing. Second, the difference between cation sizes induces local strains; individual linkers can adapt to these strains more or less efficiently, depending on their coordination environments, and the latter contribute significantly to the structure's total energy.

When both effects have comparable importance, they can either cooperate [case of MOF-5(Zn,Ca)], which amplifies both intra-cluster charge transfers and the energetic stability of specific spatial distributions; or compete[case of MOF-5(Cd,Mg)], and result in systems with generally higher mixing energies and a more complex structure-stability correlation.  The cluster coordination number is also an important factor: in UiO-66, the high cluster coordination makes linker relaxations less efficient than in MOF-5, but still of primary importance in conditioning the energies of mixed-metal structures.

However the present study only unveiled these two mechanisms; further work, both experimentally and theoretically, on a wider variety of bimetallic MOFs, would help to better understand the conditions for energetic stability. Future work will also be needed to address the question of thermodynamic stability, by accounting for both configurational and vibrational entropy in heterometallic structures. Cation (dis)ordering is also an aspect that could also be studied with a modeling at larger scale, using (i) the DFT-based energy-structure relationship to define potential energy terms and (ii) statistical energy sampling on larger systems to estimate various types of cation order parameters. More generally, one could find inspiration from studies of configurational disorder in inorganic chemistry (alloys, ...). The question of whether cations order in bimetallic MOFs could also be put in perspective with recent findings of correlated disorder in ligand-defective UiO-66(Zr) \cite{Cliffe2014, Cliffe2015}, with an underlying mechanism still only partially understood.

Finally, it is important to note that in real UiO-66 samples, one ought to take into account common defects such as frequently-occurring linker vacancies \cite{Cliffe2015}, with an average cluster coordination number that can be closer to 11 than the nominally expected value of 12. In such a situation, not all \emph{cis} linkers are equivalent, since those near a linker vacancy can rotate more easily than in the case of MOF-5. The influence of linker configurations on mixing energies could thus be less pronounced than in the fully-coordinated systems we considered in our calculations. More importantly, we suspect that cation substitutions can occur more easily at a site close to one or several linker vacancies: as exemplified in the extreme case of isolated clusters (see Table~\ref{isoclus}), the capping groups (e.g. HO$^-$, H\e{2}O, or formate) can move much more easily than the \emph{bdc} linkers, in order to adapt a cation size mismatch. From this point of view, ligand vacancies should not be avoided if one aims at large cation substitution rates (e.g. to promote or engineer catalysically-active sites). Another important aspect, not evoked here, is the mechanical stability, which is lower in systems with high rate of ligand vacancies.\cite{Thornton_Dalton16} Similarly, it may be worth investigating the conditions for mechanical stability, and more generally the mechanical properties of bimetallic MOFs.

\begin{acknowledgement}
We acknowledge access to high-performance computing platforms provided by a GENCI grant (x2016087069).
\end{acknowledgement}

\begin{suppinfo}
Figures and tables of data on bimetallic MOF-5 and UiO-66 structures.
\end{suppinfo}

\bibliography{article}

\begin{tocentry}
\includegraphics[height=47mm]{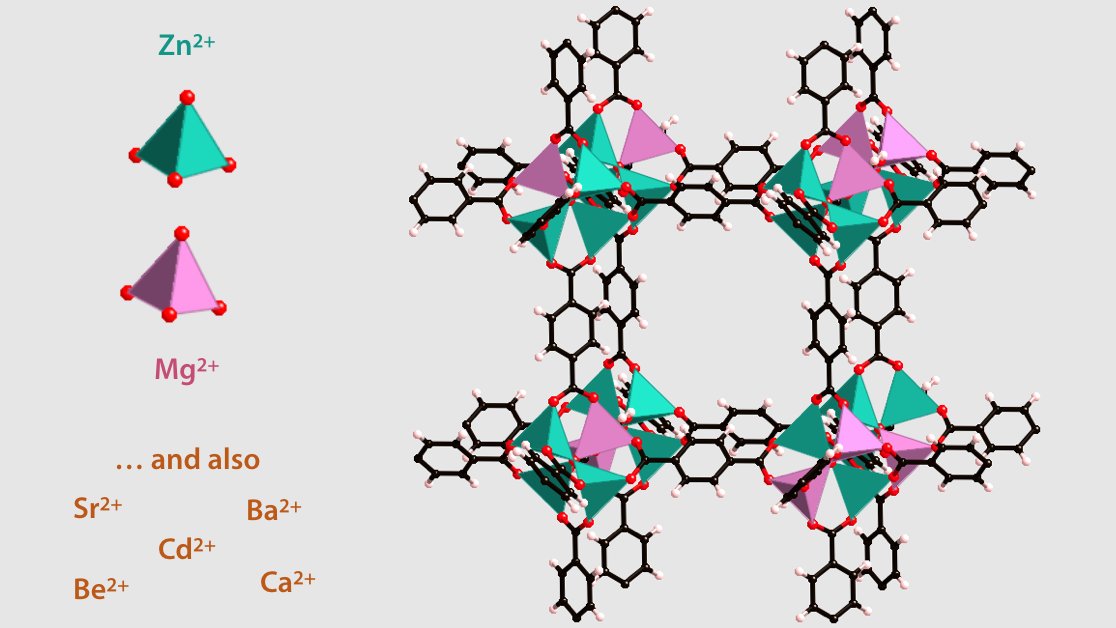}
\end{tocentry}

\end{document}